\documentclass[a4paper,12pt]{elsarticle}
\usepackage{subfigure,amssymb,amsmath}
\newcommand{\eps}{\varepsilon}
\renewcommand{\vec}[1]{\mathbf{#1}}

\begin{document}
\begin{frontmatter}
	
\title{Analyzing Near-Field Intensity Distribution\\ in Subwavelength Gratings\\ through Cylindrical Wave Decomposition}
\author[label1]{A.S. Bereza}
\affiliation[label1]{organization={Institute of Automation and Electrometry, Siberian Branch, Russian Academy of Sciences},             
addressline={1 Koptjug Avenue},            
city={Novosibirsk},
postcode={630090},
country={Russia}}
\author[label1,label2]{A.E. Chernyavsky}
\affiliation[label2]{organization={Novosibirsk State University},             
	addressline={2 Pirogov Street},            
	city={Novosibirsk},
	postcode={630090},
	country={Russia}}
\author[label3]{S.V. Perminov}
\affiliation[label3]{organization={Rzhanov Institute of Semiconductor Physics, Siberian Branch, Russian Academy of Sciences},             
	addressline={13 Lavrentiev Avenue},            
	city={Novosibirsk},
	postcode={630090},
	country={Russia}}
\author[label1]{D.A. Shapiro\fnref{fn1}}
\ead{shapiro@iae.nsk.su}
\fntext[fn1]{Corresponding author}

\begin{abstract}The investigation into the scattering of plane waves by a periodic array of parallel cylinders utilizes the method of cylindrical wave decomposition, thereby reducing the problem complexity to a series of linear algebraic equations. This methodology proves particularly efficacious when the diameter of cylinders is significantly less than the wavelength of incident wave, resulting in a rapid diminution of the solution coefficients as a function of azimuth numbers. Such a reductionist approach facilitates the computation of  scattered radiation intensity in near field. Subsequent cross-validation with numerical results corroborates the theoretical findings, showcasing a qualitative concordance between the two. This study underscores the efficacy of cylindrical wave decomposition in simplifying and accurately modeling wave scattering phenomena in structured media.
\end{abstract}
\begin{keyword}
	cylindrical wave decomposition \sep
	subwavelength gratings \sep
	near-field intensity distribution \sep
	computational electrodynamics \sep light scattering	
\end{keyword}
\end{frontmatter}

\section{Introduction}
Subwavelength gratings (SWGs), characterized by their periods being shorter than the wavelength of the interacting radiation, pose a significant challenge in the realm of nanoscale light-matter interactions \cite{fanchini2024}. Moreover, SWGs find extensive application across a broad spectrum of technological fields. They are integral to the functioning of optical filters, waveguides, and laser systems, facilitating the directed propagation of light \cite{kazanskiy2021}. Additionally, these gratings act as highly sensitive sensors capable of detecting chemical and biological substances by monitoring alterations in diffraction patterns upon analyte interaction \cite{luque2021}, and they are instrumental in the operation of high-resolution spectral devices \cite{wang2022}. SWGs are also pivotal in the synthesis of metamaterials endowed with unique optical properties, including negative refraction indices and light superconductivity \cite{cheben2023}. Their application extends to the field of lithography for the production of micro- and nanoelectronics, where they are utilized in the creation of light masks and patterns \cite{ushkov2020}, and they play a significant role in the advancement of silicon photonics \cite{wang2016}. Furthermore, SWGs contribute to the enhancement of display technologies by elevating image quality and reducing power consumption \cite{Petrov2018}, and they augment the efficiency of solar cells through increased light absorption, optimized polarization properties, and minimized reflection \cite{yuan2023,eskandari2024}. In the domain of security and anti-counterfeiting, SWGs are deployed in the fabrication of protective holograms and optical elements that present challenges in replication \cite{dalloz2022}. The myriad applications of subwavelength diffraction gratings highlight their vast potential across diverse scientific and technological fields.

In far-field analysis, subwavelength gratings (SWGs) are modeled as a uniform layer with average thickness. In specific scenarios, the effective dielectric permittivity tensor must exhibit anisotropy \cite{nemykin2022}. An alternative analytical method proposes a planar grating model within an optically homogeneous environment, abstracting grating wires as a periodic dielectric constant variation \cite{efremova2021}. Conversely, the near-field scenario presents intricacies, with detailed distributions. Scholarly discourse extensively explores cylinders, particularly the complexities of calculating the lattice sum \cite{twersky1962,lee1990,kavaklioglu2002,natarov2011}. Recent studies of near-field distribution have unveiled phenomena like nanojets, indicating localized field intensity enhancements \cite{schafer2012}, and lower mode propagation within the grating \cite{belan2015}. Despite advancements, near-field distribution understanding during scattering remains limited. This aspect has been analyzed using various numerical methods, yet not incorporating cylindrical wave decomposition (CWD) \cite{chernyavsky2023}. Our research aims to elucidate intensity distribution using CWD and rigorous coupled-wave analysis, enhancing understanding of near-field dynamics in SWGs.

Section \ref{s:theory} introduces the cylindrical wave decomposition (CWD) method and concentrates on deriving algebraic equations for the coefficients of decomposition. When considering the SWG limit, we simplify the system to derive approximate formulas for these coefficients. Section \ref{s:CWD} details the calculation of the magnetic field intensity using these derived formulas. Section \ref{s:compare} offers a comparative analysis of the results against those obtained from the Finite Element Method (FEM), perturbation theory, and the Discrete Dipole Approximation (DDA). Finally, Section \ref{s:conclusions} summarizes our findings.

\section{Decomposition over cylindrical waves}
\label{s:theory}
\begin{figure}\centering
\includegraphics[width=3in]{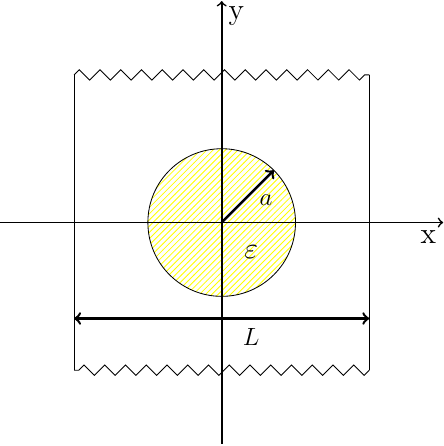}	
\caption{Unit cell geometry: section in the plane perpendicular to the cylinder axis.}\label{f:sketch}
\end{figure}

Our study addresses the diffraction problem presented by a periodic array of parallel cylinders aligned along the $z$-axis, with the cross-sectional view illustrated in Fig.~\ref{f:sketch}. We define the cylinder radius as $a$, its dielectric constant as $\eps$, and the lattice period as $L$. We select a unit cell centered on the cylinder's cross-section and investigate the scenario of normal incidence, with wave propagation along the positive $y$-axis.

For the $p$-wave, we can write the Helmholtz equation for the magnetic field $H_z(x,y)$:
\begin{equation}\label{Helmholtz}
	\bigtriangleup H_z+k^2H_z=0,
\end{equation}  
where $k=k_0\sqrt\eps$ is the wave number inside the cylinder, $k_0=\omega/c$ is the wave number outside. Here $\omega$ is the frequency of radiation, $c$ is the speed of light in free space. 

\subsection{Boundary Conditions}
The boundary conditions at the dielectric interface are simplified to the continuity of the tangential components of the electric and magnetic fields: $[E_\tau]=[H_\tau]=0$, with the square brackets indicating a discontinuity of the corresponding value inside. This results in two boundary conditions: the continuity of the magnetic field and its weighted normal derivative at the interface between the dielectric and free space at $r=a$:
\begin{equation}\label{interface}
	H_z^{in}=H_z^{out},\quad \frac{1}{\eps}\frac{\partial H_z^{in}}{\partial r}=\frac{\partial H_z^{out}}{\partial r},
\end{equation}
where $r=\sqrt{x^2+y^2}$ is the radial coordinate, ``in'' indicates the interior region of the circle, and ``out'' refers to the exterior region. For the periodic structure problem, the periodicity condition on the cell's sides complements Eq. (\ref{interface}):
\begin{equation}\label{periodic}
	H_z(-L/2,y)=H_z(L/2,y).
\end{equation}
In cases of oblique incidence, the phase term $e^{ik_0xL}$ is introduced in condition (\ref{periodic}) as per the Floquet theorem.

\subsection{Fourier Series}
The magnetic field in the observation point can be represented as the sum of incident and scattered wave $H_z^s$
\begin{equation}\label{scattered}
	H_z(\vec r)=e^{ik_0y}+H_z^s(\vec r).
\end{equation}
The incident plane wave in turn can be decomposed into a Fourier series in azimuth angle $\varphi$: 
\begin{equation}\label{plane-wave}
	e^{ik_0y}=e^{ik_0r\sin\varphi}=\sum_{m=-\infty}^\infty
	e^{im\varphi}J_m(k_0r),\quad r>a.
\end{equation}

The field inside $n$-th cylinder can also be written as an angular harmonic expansion:
\begin{equation}\label{Bessels}
	H_{z,n}^{in}(\vec r)=\sum_{m=-\infty}^{\infty}
	a_{m}J_m\left(k|\vec r-\vec r_n|\right)e^{im\varphi_n},\;
	|\vec r-\vec r_n|<a,\;\varphi_n=\arg(\vec r-\vec r_n).
\end{equation}
Here $\vec r_n=(nL,0)$ are coordinates of the $n$th cylinder center. The scattered wave can be written as the sum of waves from each cylinder:
\begin{align}
	H_z^s(\vec r)=\sum_{n=-\infty}^\infty\sum_{m=-\infty}^\infty
	b_{m}H_m^{(1)}\left(k_0|\vec r-\vec r_n|\right) e^{im\varphi_n},\quad|\vec r-\vec r_n|>a.\label{partial}
\end{align}
Hereafter $\vec r=x+iy$ is 2-dimensional radius-vector.  The coefficients of inner and outer series $a_m,b_m$ are independent of the cylinder number, as follows from periodicity (\ref{periodic}) and a specific excitation source in the form of a plane wave with normal incidence. The obtained Fourier series (\ref{partial}) is the required decomposition into cylindrical waves.

\subsection{Equations for Coefficients}
One can simplify the series exploiting the Graf addition theorem \cite{Olver10} 
\begin{equation}\label{addition}
	e^{i\nu\chi}Z_\nu^{(1)}(w)=\sum_{k=-\infty}^\infty
	e^{ik\alpha}Z_{\nu+k}^{(1)}(u)J_k(v),
\end{equation}	
where $Z_\nu$ is an arbitrary cylindrical function.
Here, the vectors corresponding to the complex numbers $u,v,w$ form a triangle in the complex plane, with $v$ being the smallest of its sides, $\alpha,\chi$ are the interior angles of this triangle, opposite to sides $v,w$, respectively. It yields the Hankel function as a series of Bessel functions. Using Eq. (\ref{addition}) we get
\begin{equation}\label{Graf}
	e^{il\varphi_n}H_{l}^{(1)}(k_0|\vec r-\vec r_n|)=	
	\sum\limits_{m=-\infty}^\infty
	e^{im\varphi}H_{(m-l)\sigma}^{(1)}(k_0r_n)J_m(k_0r),
\end{equation}
where $\sigma=\mbox{sign}\,n$. 

Substituting (\ref{Graf}) into (\ref{partial}) we find the scattered field
\begin{equation}
	H_z^s(\vec r)=\sum_{l=-\infty}^\infty b_{l}\left(
	\sum_{m=-\infty}^\infty J_m(k_0r)
	e^{im\varphi}F_{ml}+ e^{il\varphi}H_{l}^{(1)}(k_0r)\right).
	\label{double_sum}
\end{equation}
Here
\begin{equation}
	F_{ml}=\sum_{n\neq0}H_{(m-l)\sigma}^{(1)}(k_0r_n)
	\label{Fml}
\end{equation}
is the lattice sum. Matrix element $F_{ml}$ is a superposition of cylindrical waves (Hankel functions) scattered by all the cylinders except of $n=0$:
\begin{equation}\label{BV-notation}
	F_{ml}=\left[\sum_{n=1}^{\infty}+\sum_{n=-\infty}^{-1}\right]
	H_{m-l}^{(1)}(k_0|n|L)=\sum_{n=1}^\infty H_{m-l}^{(1)}(k_0nL)\left(1+(-1)^{m-l}\right),
\end{equation}
hence $F_{-m,-l}=F_{ml}$. The matrix elements $F_{ml}$ depend only on the difference of indices, i.e. the matrix is Teplitzian. It opens the possibility of fast algorithm of its inversion \cite{blahut2010}.

From the inner expansion (\ref{Bessels}), the outer expansion (\ref{double_sum}) and the boundary conditions (\ref{interface}),
equating the coefficients at equal angular harmonics $m$, we find the algebraic system for the coefficients:
\begin{align}
	J_m+J_m\sum_lF_{ml}b_l+H_mb_m=a_mL_m,\label{short1}\\
	J_m'+J_m'\sum_lF_{ml}b_l+H_m'b_m=\frac{a_m}{\sqrt\eps}L_m^{\prime},\label{short2}
\end{align}
where $J_m\equiv J_m(k_0a),H_m\equiv H_m^{(1)}(k_0a), L_m\equiv J_m(ka)$. 

\begin{figure} \centering
	\includegraphics[width=4in]{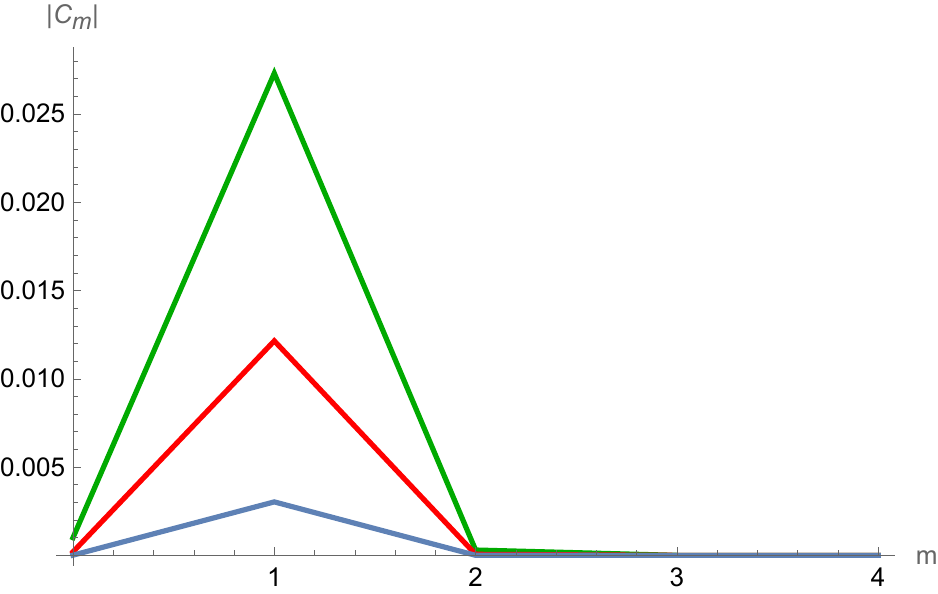}
	\caption{Coefficients $|C_m|$ at $\eps=2.25$ and different $m=0,\dots,4$. From top to bottom $k_0a=0.3$, 0.2, 0.1.}
	\label{f:coefficients}
\end{figure}

We find the inner coefficient $a_m$ from (\ref{short1}), substitute them into (\ref{short2}), and get a system on the outer coefficients $b_m$:
\begin{equation}\label{system}
	b_m+C_m\sum_{l}F_{ml}b_l+C_m=0,
\end{equation}	
where
\begin{equation}\label{Wronskians}	
	C_m=\frac{J_mL_m'/\sqrt\eps- J_m'L_m}{H_mL_m'/\sqrt\eps- H_m'L_m}
\end{equation}	
are auxiliary values, the coefficients of decomposition for a single isolated cylinder \cite{Harrington61}.
Azimuth number $m=\pm1$ corresponds to dipole radiation; $m=0$ is responsible for isotropic scattering. The other terms of the expansion are small if the phase runup on the characteristic transverse dimension of the cylinder is small $ka\ll1$. Fig. \ref{f:coefficients} confirms that the dipole scattering is the most significant for $p$-wave. 

\section{Intensity Distribution}\label{s:CWD}
Here we find the near-field intensity distribution of the scattered wave with CWD. In the next section we compare our pattern with the results of calculation by other methods.

The scattered field in the external region is found from Eq (\ref{double_sum}):
\begin{align}
	H_z^s(\vec r)=\sum_{n=-N_{\max}}^{N_{\max}}
	\sum_{m=-M_{\max}}^{M_{\max}}
	b_me^{im\varphi}H_m^{(1)}(k_0|\vec r-\vec r_n|).
	\label{raschet}	
\end{align}	
Coefficients $b_m$ are determined from the truncated system (\ref{system}), with the summation constrained to $|m|\leqslant M_{\max}$ and $|n|\leqslant N_{\max}$. In our calculations, we set $M_{\max} = 1$ and $N_{\max} = 50$. The results indicate that the contribution of $m = 2$ is two orders of magnitude smaller than that of $m = 1$. To accommodate  increased geometric filling factor $2a/L\leqslant1$, a greater number of wires, $2N_{\max} + 1$, must be included in the calculations.

\begin{figure}\centering
	\subfigure[]{\includegraphics[width=4.5in]{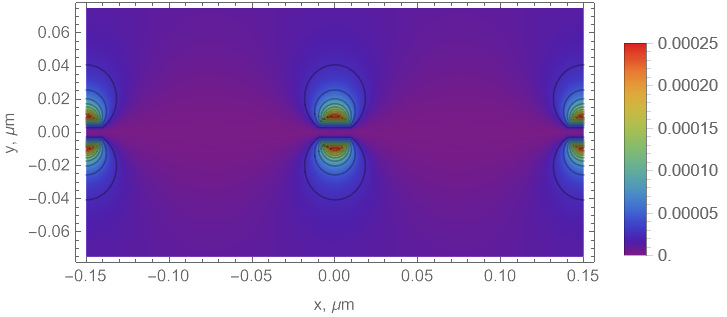}}
	\subfigure[]{\includegraphics[width=4.5in]{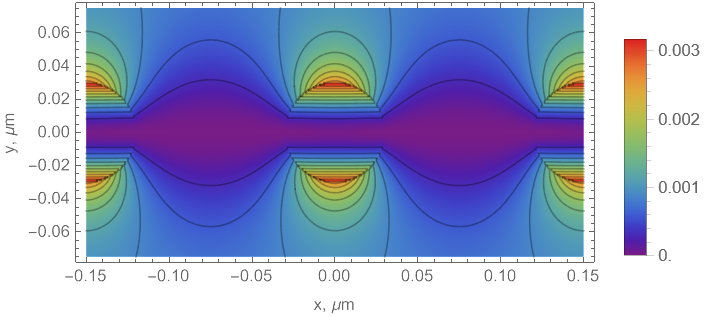}}
	\subfigure[]{\includegraphics[width=4.5in]{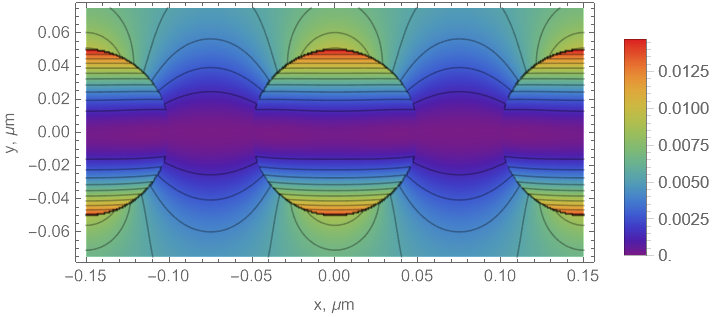}}
	\caption{Intensity distribution $|H_z(x,y)|^2$ at wavelength $\lambda=1.512~\mu$m, $N_{\max}=50, M_{\max}=1, \eps=2.25, L=0.15~\mu$m: $a=0.01$  (a), $0.03$ (b), $0.05$ (c).
	}\label{f:CWD}
\end{figure}	

The distribution, as depicted in Fig. ref{f:CWD}, is illustrated for a constant period $L = 0.15mu$m across varying cylinder radii: $a = 0.01, 0.03, 0.05mu$m, thereby modifying the filling factor. Distinct features of this distribution include the intensity $|Hz|^2$ peaking at the cylinder's top and bottom edges, notably at the interface between free space and the dielectric at $x = nL, y = pm a$. The peak's amplitude escalates by an order of magnitude from one subfigure to the next, correlating with an increase in the filling factor. The peaks manifest as crescents, which shorten as the grating becomes sparser. Near $y = 0$, a trough is observed, expanding at the midpoint of the gap between adjacent cylinders, with its width enlarging as the filling factor decreases. This widening gap is associated with a diminution in maximum intensity. The parameters $k0a = 0.04, 0.12, 0.2$ suggest that for each cylinder configuration in Fig. ref{f:CWD}(a), a sparse grating results in a pattern similar to that of a single isolated cylinder cite{Harrington61}. As the filling factor increases in Fig. ref{f:CWD}(c), the interaction between cylinders becomes more pronounced.

\section{Comparing with other methods}\label{s:compare}

\begin{figure}\centering
	\subfigure[]{\includegraphics[width=4.5in]{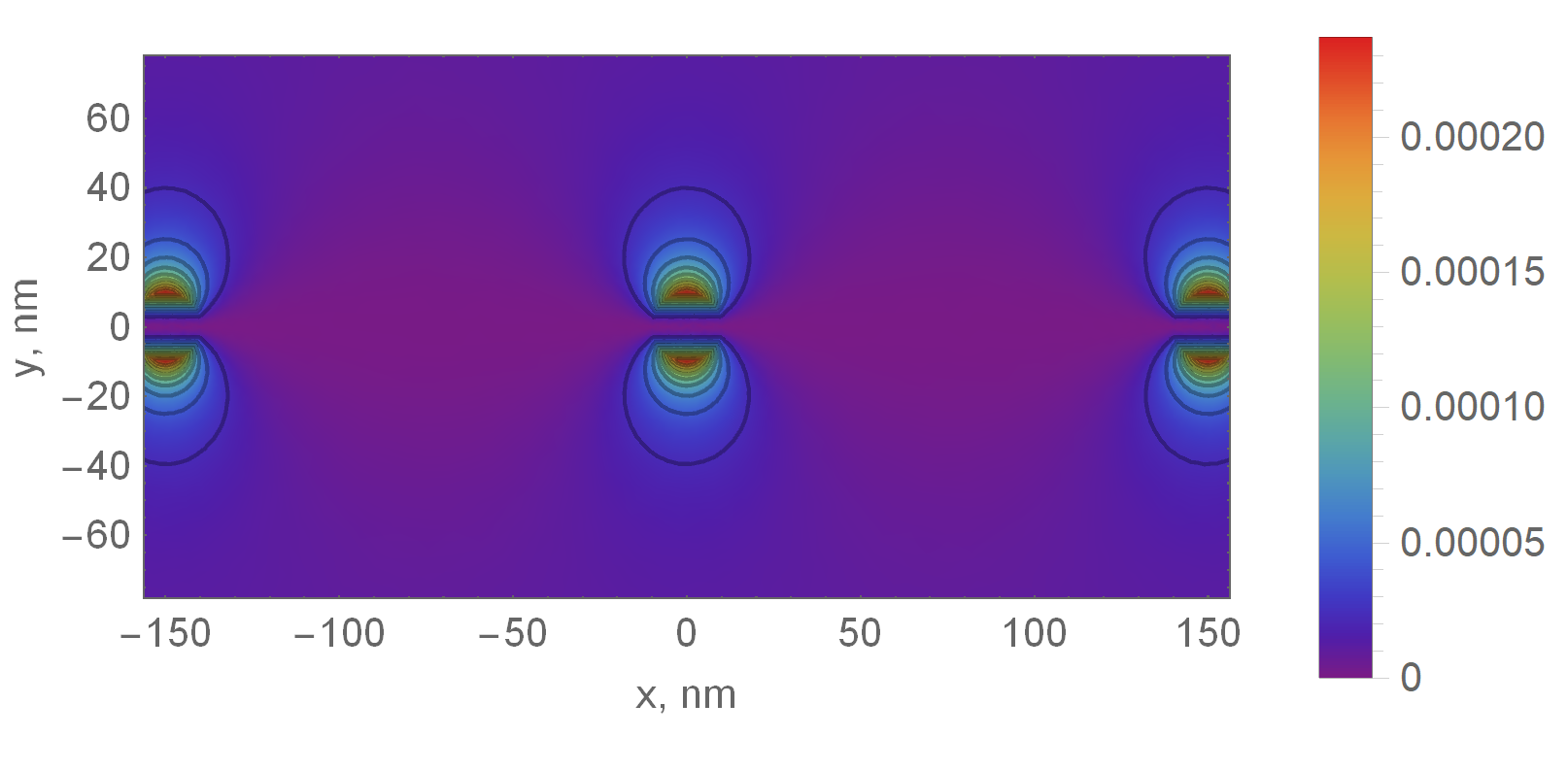}}
	\subfigure[]{\includegraphics[width=4.5in]{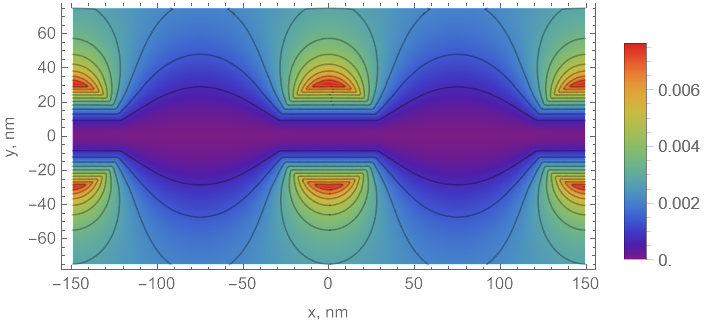}}
	\subfigure[]{\includegraphics[width=4.5in]{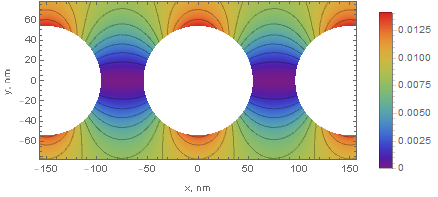}}
	\caption{Intensity distribution $|H_z(x,y)|^2$ calculated at the same parameters, as in Fig. \ref{f:CWD}(a,b,c): finite elements method (a),  perturbation theory (b), discrete dipole approximation (c).}\label{f:numerical}
\end{figure}	

\subsection{Finite Elements Method}
To compare the results, we used the COMSOL Multiphysics 6.1 environment \cite{COMSOL22} with the Wave Optics package. Geometry is defined as rectangle domain with wide $20\; \mu$m and height $1.8\; \mu$m containing a chain of cylinders. The domain material is air with $\eps=1$, surrounded by layers of depth $d=0.3 \;\mu$m where the perfectly matched layers boundary condition is specified. Chain is linear array of 100 identical cylinders (dielectric constant $\eps=2.25$) displaced along $x$-axis with respect to parameters of problem: period $L=0.15\;\mu$m between the centers, radii are varying $a=0.01,0.03,0.05\; \mu$m. Background field defined as plane wave expansion of Gaussian beam is incident along $y$ axis: width $w_0= 7.5\; \mu$m, focal plane at $y=0$, wavelength $\lambda=1.512\; \mu$m. We use inhomogeneous mesh of triangles ranging in size from 1 nm inside and in the immediate vicinity of the cylinder to 30 nm in the outer region.

\subsection{Perturbation Theory}
Due to the normal incidence of the external field on the periodic lattice of cylinders $\eps(x,y)=\eps(x+L,y)$, the scattered field will also be a periodic function $E^s(x,y)=E^s(x+L,y)$, which allows us to represent $E^s(x,y)$ as a Fourier series \cite{chernyavsky2023}:
\begin{equation}\label{Fourier}
	E_x(x,y)=\sum_{n=-\infty}^{\infty}U_n(x)e^{2\pi inx/L},\quad
	E_y(x,y)=\sum_{n=-\infty}^{\infty}V_n(x)e^{2\pi inx/L}.	
\end{equation}
For $p$-wave we substitute series (\ref{Fourier}) into Maxwell's equation for the electric field
\begin{equation}\label{Maxwell}
	\nabla\times\nabla\times
	\mbox{\mathversion{bold}$E$}=\eps(\vec r)k_0^2\mbox{\mathversion{bold}$E$}.
\end{equation}
Equations for components are:
\begin{equation}\label{components}
	(E_y)_{xy}-(E_x)_{yy}=\eps(\vec r)k_0^2E_x,\quad
	(E_y)_{xx}+(E_x)_{xy}=\eps(\vec r)k_0^2E_y.
\end{equation}

Considering the right side of equations (\ref{Maxwell}), (\ref{components}) as a small perturbation, assuming the external field equal to $E_x^{(0)}=U_0^{(0)}=e^{ik_0y}, E_y^{(0)}=0$, we obtain
\begin{align}
	&V_0^{(1)}=0,\quad
	V_n^{(1)}=-\frac{2\pi in}{Lq_n^2}\frac{dU_n^{(1)}}{dy},\quad
	U_0^{(1)}=\frac{i(\eps-1)k_0}{2L}\pi a^2e^{-ik_0y}\nonumber\\&-(\eps-1)k_0\int_{-a}^a\theta(y-\Tilde{y})
	\sin\left(k_0(y-\Tilde{y})\right)
	\frac{2\sqrt{a^2-\Tilde{y}^2}}{L}
	e^{ik_0\Tilde{y}}\,d\Tilde{y},\nonumber\\
	&U_n^{(1)}=-(\eps-1)q_n^2\int_{-a}^a
	\frac{e^{-q_n|y-\Tilde{y}|}}{2q_n}\frac{e^{ik_0\Tilde{y}}}{\pi n}
	\sin\frac{2\pi n\sqrt{a^2-\Tilde{y}^2}}{L}\,d\Tilde{y}.&
	\label{orders}
\end{align}	
Here $q_n^2=(4\pi n)^2/L^2 -k_0^2$ is the Fourier parameter of mode $n$, $V_n^{(1)}$ and $U_n^{(1)}$ are the principal terms of the perturbation ($V_n^{(m)}$ and $U_n^{(m)}$ are the $m$-th terms, respectively), $\theta$ is the Heaviside step function. If the scattered field being small, it is sufficient to retain $V_n^{(1)}$ and $U_n^{(1)}.$

Knowing $E_x^{((1))}$ and $E_y^{((1))}$, we can find  magnetic field $H_z^{((1))}$:
\begin{equation}
	H_z^{(1)}=\frac{1}{ik_0}\left[(E_y)_x-(E_x)_y\right]=
	\frac{1}{ik_0}\sum_{n\neq0}k_0^2
	\frac{U_n^{(1)\prime}(y)}{q_n^2}-
	\frac{U_0^{(1)\prime}(y)}{ik_0}.\label{magnetic-field}
\end{equation}
In contrast to expansion (\ref{Fourier}), series (\ref{magnetic-field}) converges faster due to the continuity of magnetic field in space (the electric field is piecewise continuous, discontinuities on the cylinder boundary worsen convergence). We hold $|n|\leqslant N_{\max}=30$ terms in Eq. (\ref{magnetic-field}), considering 61 terms of the series.

\subsection{Discrete Dipole Approximation}
The Discrete Dipole Approximation (DDA) method, extensively reviewed in \cite{Yurkin07} and originally introduced in \cite{purcell73,Draine88}, has been a staple analytical tool in nanophotonics, aerosol and hydrosol optics, and astrophysics for decades. This method involves discretizing the scattering object into small cells treated as point dipoles, followed by solving the system of integral equations that describe dipole interactions. DDA is particularly adept at determining scattered fields and analyzing their dependence on incidence angle and wavelength.

While predominantly applied to three-dimensional problems, DDA has also shown efficacy in two-dimensional contexts, both in uniform \cite{Martin98,olPFS19} and compound environments \cite{plaPS22}. In this study, we employed 2-dimensional DDA to compute the scattered magnetic field from a finite chain of 49 cylinders excited by an incident plane wave. Each cylinder's cross-section was divided into 352 square cells, equivalent to two-dimensional dipoles. By solving the equations for coupled dipoles, as detailed in \cite{olPFS19}, we obtained a total of 17,248 dipoles. These facilitated the calculation of scattering field in regions external to, and slightly distant from, the cylinder surfaces, due to the inaccuracy of dipole-emitted field calculations near edges.

\subsection{Comparison}
The intensity distribution derived from various computational methods --- FEM, perturbation series, and DDA --- is compared, utilizing the same parameters as for the CWD. As depicted in Fig. \ref{f:numerical}, analogous to Fig. \ref{f:CWD}, notable intensity peaks are observed at the cylinder tops and bottoms, alongside a relatively narrow trough along the $x$-axis at small $|y|$ values.

Fig. \ref{f:numerical}(a) illustrates results obtained through FEM within the COMSOL Multiphysics framework, revealing intensity peaks at the cylinder extremities and diminished value within the substantial gap between cylinders, akin to observations in Fig. \ref{f:CWD}(a). Fig. \ref{f:numerical}(b), derived using perturbation theory, exhibits an intensity increase correlating with the filling factor, mirroring the pattern seen in Fig. \ref{f:CWD}(b). Fig. \ref{f:numerical}(c) showcases the results from DDA, where white circles indicate uncalculated fields inside and immediately outside the cylinders, yet the external pattern closely aligns with the CWD results in Fig. \ref{f:CWD}(c).

Therefore, the characteristic features identified in CWD analyses are corroborated by findings from established numerical methods, affirming qualitative consistency across gratings with diverse filling factors.

\section{Conclusions}\label{s:conclusions}

The CWD method yields accurate near-field intensity distributions, consistent with computational electrodynamics and perturbation theory. Key characteristics of the distribution include:
(i) Intensity peaks at coordinates $x=nL, y=\pm a$, where $n$ represents integers;
(ii) A narrow trough near the $x$-axis, broadening with distance from between cylinders. In summary, CWD proficiently analyzes wave scattering in cylindrically symmetric systems, offering detailed near-field and modal insights.

\section*{Acknowledgments} The authors thank O.V. Belai  A.V. Nemykin, and L.L. Frumin for helpful discussions. The work was funded by the Russian Science Foundation, grant \#24-22-00087.


\end{document}